\documentstyle[12pt]{article}
\newlength{\extraspace}
\newlength{\extraspaces}
\newcommand{\ba}{\begin{eqnarray}
\addtolength{\abovedisplayskip}{\extraspaces}
\addtolength{\belowdisplayskip}{\extraspaces}
\addtolength{\abovedisplayshortskip}{\extraspace}
\addtolength{\belowdisplayshortskip}{\extraspace}}
\newcommand{\ea}{\end{eqnarray}}
\begin{document}

\thispagestyle{empty}

\hfill \parbox{3.5cm}{hep-th/ \\ SIT-LP-05/12}
\vspace*{1cm}
\vspace*{1cm}
\begin{center}
{\large \bf Cosmology  \\ 
and                    \\ 
Nonlinear Supersymmetric General Relativity}  
\footnote{Based on the talk given by K. Shima at International Europhysics Conference on High Energy Physics, 
July 21st-27th, 2005, Lisboa, Portugal}    
\\[20mm]
{\sc Kazunari Shima}${}^{\rm a}$\footnote{
\tt e-mail: shima@sit.ac.jp}, \
and \
{\sc Motomu Tsuda}${}^{\rm a}$\footnote{
\tt e-mail: tsuda@sit.ac.jp}
\\[5mm]
${}^{\rm a}${\it Laboratory of Physics, 
Saitama Institute of Technology \\
Okabe-machi, Saitama 369-0293, Japan} \\[3mm]
{December 2005}\\[15mm]


\begin{abstract}

We discuss cosmological implications of nonlinear supersymmetric(NLSUSY) 
general relativity(GR) of the form of Einstein-Hilbert(EH) action for empty spacetime, 
where NLSUSY GR is obtained by the geomtrical arguments  on new spacetime 
just inspired by NLSUSY. 
The new action of NLSUSY GR is unstable and breaks down spontaneously to 
EH action with Nambu-Goldstone(NG) fermion matter.
We show that NLSUSY GR elucidates the physical meanings of the cosmologically important quantities, 
e.g., the spontaneous SUSY breaking scale, the cosmological constant, the dark energy and the neutrino mass 
and describe natually the paradigm of the accelerated expansion of the present 
universe. 

\end{abstract}
\end{center}

\newpage
\section{Introduction}

SUSY\cite{wz}\cite{va}\cite{gl} and its spontaneous breakdown\cite{fi}\cite{o} are conformed to the rationale of spacetime, therefore 
necessary to be accessed not only from  the low energy particle physics but also from the cosmology. 
Despite the success of the standard model(SM), 
facing so many arbitrary paramerters and mysterious textural features, 
e.g. the CP violation phase, mixing angles, generation structures, 
tiny neutrino masses, etc., 
we are tempted to suppose that they may be certain composites  and/or that 
they should be attributed to the  particular geometrical structure of spacetime 
which unifies the (classical) notions: the object  and  
the background space-time manifold irrespective of being classical and quantum.   \\
From the simplicity and the beauty of nature, 
the NLSUSY\cite{va} invariant coupling of graviton with the spin ${1 \over 2}$ fermion is 
anticipated for materializing the standard model(SM) with the spontaneous 
breakdown of SUSY\cite{ks1}. \\
We  have shown group theoretically in ref.\cite{ks1} 
that  among the massless irreducible representations 
of all SO(N) super-Poincar\'e(SP) groups 
N=10 SP group is the only one that contains the SM 
with just three genarations of quarks and leptons, where by extending the Gell-Mann's  idea\cite{mg} 
we have decomposed 10 spinor supercharges into  
${\underline 10 = \underline 5 + \underline 5^{*}}$ with respect to SU(5). 
Surprisingly, the fundamental representation $\underline 5$ has the same quantum number 
as the  $\underline 5$ of observed quarks-leptons multiplet of SU(5) grand unified theory(GUT)\cite{gg}, 
i.e. ${{\underline 5}_{SGM}={\underline 5}_{GUT}}$. 
Regarding 10 {\it supercharges} as the hypothetical fudamental spin 1/2 {\it particles(superons)}-quintet 
and anti-quintet, we have proposed the composite superon-graviton model(SGM) for nature\cite{ks2}\cite{st1}. 
In SGM, all (observed) elementary  particles except graviton are assumed 
to be superon(supercharge)-composites  
and are {\em eigenstates}  of SO(10) SP symmetry in the sence  that as shown partly in Sec. 3 
each state of superon(supercharge)-composites  can be recasted algebraiclly 
into the equivalent local field of the  supermultiplet of the corresponding 
linear(L) SUSY explicitly.  
We have attempted to reproduce the Feynmann diagramms of  SU(5) GUT in SGM picture  by replacing 
the sigle-line of the propagator of a particle into the multiple-lines of the constituents of 
each particle\cite{sts}.  
Despite SU(5) GUT-like features  ${ {\underline 5}_{SGM}={\underline 5}_{GUT}}$,  
the SUSY-composite picture of SGM, 
i.e. the constituent(superon) number conservation at the vertex, gives new remarkable insights 
into the proton decay, R-parity, neutrino oscillations, $\cdots$, etc. 
These group theoretical arguments suggest the NLSUSY invariant coupling of graviton 
with the spin ${1 \over 2}$ fermion(superon) and the {\it spin ${1 \over 2}$ field-current(supercharge) 
identity}, i.e. N=10 NLSUSY Volkov-Akulov(VA) model~\cite{va} coupled with gravity. 
SUSY and its spontaneous breakdown can be encoded in a self-contained way. 

\section{Nonlinear Supersymmetric General Relativity \\
(NLSUSY GR)}
The curvature of particular spacetime, the ultimate physical entity, 
materializes the general relativity of nature. 
We extend the geometrical arguments of Einstein general relativity(EGR) on Riemann spacetime 
to new  (called {\it SGM}  hereafter)  spacetime where the tangent spacetime posesses 
{\it local NLSUSY degrees of freedom(d.o.f)}, 
i.e. besides the ordinary SO(3,1) Minkowski coordinate $x^{a}$ 
the SL(2C) Grassman coordinates $\psi$ for the coset space ${superGL(4,R) \over GL(4,R)}$ 
turning subsequently to the NG fermion (called {\it superon} hereafter) dynamical d.o.f. 
are attached at every curved spacetime point. 
(Note that locally homomorphic SO(3,1) and SL(2C) are  non-compact groups for 
spacetime d.o.f which is analogous to  SO(3) and SU(2)  compact groups for  gauge d.o.f. 
of 't Hooft-Polyakov monopole\cite{th}\cite{p}.) 
We have obtained  the following NLSUSY GR(N=1 SGM) action\cite{ks2} of the vacuum EH-type in SGM spacetime.%
\begin{equation}
L(w)=-{c^{4} \over 16{\pi}G}\vert w \vert(\Omega + \Lambda ),
\label{SGM}
\end{equation}
\begin{equation}
\vert w \vert=det{w^{a}}_{\mu}=det({e^{a}}_{\mu}+ {t^{a}}_{\mu}(\psi)),  \quad
{t^{a}}_{\mu}(\psi)={\kappa^{2}  \over 2i}(\bar{\psi}\gamma^{a}
\partial_{\mu}{\psi}
- \partial_{\mu}{\bar{\psi}}\gamma^{a}{\psi}),
\label{w}
\end{equation} 
where $w^{a}{_\mu}(x)$ invertible and $\Omega$ are the unified vierbein 
and the the unified scalar curvature of SGM spacetime, respectively. 
Accordingly $s_{\mu\nu} \equiv w^{a}{_\mu}\eta_{ab}w^{b}{_\nu}$ and 
$s^{\mu \nu}(x) \equiv w^{\mu}{_{a}}(x) w^{{\nu}{a}}(x)$ 
are unified metric tensors of SGM spacetime\cite{ks2}\cite{st1}. 
The explicit  expression of $\Omega$  is obtained  by just replacing ${e^{a}}_{\mu}(x)$  
by ${w^{a}}_{\mu}(x)$ in Ricci scalar $R$ of EGR~\cite{st1}.   
${e^{a}}_{\mu}$ describing the local SO(3,1) and 
${t^{a}}_{\mu}(\psi)$ describing the localSL(2C) are the ordinary vierbein of EGR 
and the mimic vierbein of the stress-energy-momentum tensor 
of superons $\psi(x)$, respectively.
$G$ is the Newton gravitational constant, $\Lambda$ is a  ({\it small}) cosmological term 
and ${\kappa}$ of NLSUSY VA model~\cite{va} with the dimension $(length)^{4}$ 
is now related to ${\kappa^{2} = ({c^{4}\Lambda \over 8{\pi}G}})^{-1}$. 
It is remarkable that the constant ${\kappa}$, i.e. 
the strength of the superon-vacuum coupling constant in the (low energy theorem) 
particle physics viewpoint, is determined by the quantities of spacetime,  
$G$ and $\Lambda$,  in the NLSUSY GR(SGM) scenario. 
SGM contains hierarchic two mass scales, 
 ${1 \over G}$(Planck scale) in the first term and $\kappa^{2} \sim {\Lambda \over G}$ 
in the second term describing the curvature energy and the vacuum energy of SGM, respectively.

NLSUSY GR action  (\ref{SGM}) is invariant  under the new NLSUSY transformation\cite{st2};
\begin{equation}
\delta_{\zeta}{\psi}={1 \over \kappa} \zeta - 
i\kappa (\bar{\zeta}{\gamma}^{\rho}\psi) \partial_{\rho}\psi,
\quad
\delta_{\zeta}{e^{a}}_{\mu}= i \kappa (\bar{\zeta}{\gamma}^{\rho}\psi)\partial_{[\rho} {e^{a}}_{\mu]},
\label{newsusy}
\end{equation} 
consequently GL(4R) transformations on ${{w^{a}}_{\mu}}$  
\begin{equation}
\delta_{\zeta} {w^{a}}_{\mu} = \xi^{\nu} \partial_{\nu}{w^{a}}_{\mu} + \partial_{\mu} \xi^{\nu} {w^{a}}_{\nu}, 
\quad
\delta_{\zeta} s_{\mu\nu} = \xi^{\kappa} \partial_{\kappa}s_{\mu\nu} +  
\partial_{\mu} \xi^{\kappa} s_{\kappa\nu} 
+ \partial_{\nu} \xi^{\kappa} s_{\mu\kappa}, 
\label{newgl4r}
\end{equation} 
where $\zeta$ is a constant spinor, $\partial_{[\rho} {e^{a}}_{\mu]} = 
\partial_{\rho}{e^{a}}_{\mu}-\partial_{\mu}{e^{a}}_{\rho}$,
and $\xi^{\rho}=i \kappa^{2} (\bar{\zeta}{\gamma}^{\rho}\psi)$. 
The commutators of two new NLSUSY transformations (\ref{newsusy})  on $\psi$ and  ${e^{a}}_{\mu}$ 
are GL(4R), i.e. new NLSUSY (\ref{newsusy}) is the square-root of GL(4R); 
\begin{equation}
[\delta_{\zeta_1}, \delta_{\zeta_2}] \psi
= \Xi^{\mu} \partial_{\mu} \psi,
\quad
[\delta_{\zeta_1}, \delta_{\zeta_2}] e{^a}_{\mu}
= \Xi^{\rho} \partial_{\rho} e{^a}_{\mu}
+ e{^a}_{\rho} \partial_{\mu} \Xi^{\rho},
\label{com1/2-e}
\end{equation}
where $\Xi^{\mu} = 2i\kappa (\bar{\zeta}_2 \gamma^{\mu} \zeta_1)
      - \xi_1^{\rho} \xi_2^{\sigma} e{_a}^{\mu}
      (\partial_{[\rho} e{^a}_{\sigma]})$.
They show the closure of the algebra.      \\
NLSUSY GR action (\ref{SGM}) is invariant at least under the following spacetime symmetries\cite{st1} which is locally isomorophic to super-Poincar\'e(SP): 
\begin{equation}
[{\rm new \ NLSUSY}] \otimes [{\rm local\ GL(4,R)}] 
\otimes [{\rm local\ Lorentz}]\otimes [{\rm local \ spinor \  translation(LST)}]  
\label{sgmspsymm}
\end{equation}
and the following internal symmetries for N-superons $\psi{_j}, (j=1,2,..N)$ extended NLSUSY GR: 
\begin{equation}
[{\rm global \ SO(N)}] \otimes [\rm local \ U(1)^{N}].  
\label{sgmisymm}
\end{equation}  \par
SGM action $L(w)$ (\ref{SGM}) on SGM spacetime is unstable due to the global NLSUSY structure(SP d.o.f.)      
of tangent spacetime and breakes down spontaneously to EH action  with the superon(NG fermion) matter 
in Riemann spacetime as follows 
\begin{equation}
L(e.\psi)=-{c^{4} \over 16{\pi}G}\vert e \vert \{ R(e) + \Lambda + \tilde T(e, \psi) \}, 
\label{SGMR}
\end{equation}
where $\tilde T(e,\psi)$ is the kinetic term and the gravitational interaction of superon. 
The second and the third terms produces N-etended NLSUSY VA action 
with  ${\kappa^{2} = ({c^{4}\Lambda \over 8{\pi}G}})^{-1}$
in Riemann-flat($e{^a}_{\mu}(x) \rightarrow {\delta}{^a}_{\mu}$) spacetime. 
Finally we just mention that by the straightforward extention of SGM argument 
we can obtain NLSUSY GR with spin ${3 \over 2}$ NG fermion\cite{st4} of Baaklini action\cite{b}.

\section{Linearization of NL SUSY}
SGM action (\ref{SGMR}) describing the geometry of the gravitational interactions of superons 
in Riemann spacetime is highly nonlinear. 
It is necessary to linearize SGM action (\ref{SGMR}) to obtain an equivalent and renormalizable field theory, 
where NLSUSY is recasted to the broken LSUSY defined on the LSUSY supermultiplet fields 
irreducible representations of SO(N) SP. 
N=1 VA model is linearized by many authors.\cite{ik}\cite{r}\cite{uz}
\cite{sw}\cite{stt1}
As a preliminary for SGM linearization we have shown\cite{stt2} in flat spacetime that 
N=2 NLSUSY VA model (expanded in $\kappa$): 
\begin{equation}
L_{\rm VA} =- {1 \over {2 \kappa^2}} \left[ 1 + \kappa^2 t{^a}_a 
+ {1 \over 2} \kappa^4 (t{^a}_a t{^b}_b 
- t{^a}_b t{^b}_a) \right. 
\left. - {1 \over 3!} \kappa^6 \epsilon_{abcd} \epsilon^{efgd} 
t{^a}_e t{^b}_f t{^c}_g  + \cdots
\right], 
\label{vaactex}
\end{equation}
%
which is invariant under N=2 NLSUSY transformation \\
${\delta_Q \psi_L^i = {1 \over \kappa} \zeta_L^i 
- i \kappa \left( \bar\zeta_L \gamma^a \psi_L  
- \bar\zeta_R \gamma^a \psi_R \right) 
\partial_a \psi_L^i}$, 
is equivalent to the following free action of the spontaneouly broken N=2 LSUSY: \\
\begin{equation}
L_{\rm lin} =  \partial_a \phi \partial^a \phi^* 
- {1 \over 4} F^2_{ab} 
+ i \bar\lambda_{Ri} \!\!\not\!\partial \lambda_{Ri} 
+ {1 \over 2} (D^I)^2 - {1 \over \kappa} \xi^I D^I, 
\label{lact}
\end{equation}
where $\psi_{Li},(i=1,2)$ is superon field,   
${
w{^a}_b = \delta^a_b + \kappa^2 t{^a}_b, 
t{^a}_b = - i \bar\psi_L \gamma^a \partial_b \psi_L 
+ i \bar\psi_R \gamma^a \partial_b \psi_R }$, 
$\psi_{Ri} = C \bar\psi_{Li}^T$. 
$\xi^I,(I=1,2,3)$ are arbitrary real parameters of the induced global SU(2)(SO(3)) 
rotation $(\xi^I)^2 = 1$. 
The last term is the Fayet-Iliopoulos $D$ term 
indicating spontaneous SUSY breaking with the vacuum expectation value 
$D^I = \xi^I / \kappa$. 
In these arguments all fields of LSUSY supermultiplet are 
the composites of superons  $\psi_L^i$, e.g., 
${\phi(\psi) =  {1 \over \sqrt{2}} \, i \kappa \xi^I 
\bar\psi_R \sigma^I \psi_L  
- \sqrt{2} \kappa^3 \xi^I \bar\psi_L \gamma^a \psi_L 
\bar\psi_R \sigma^I \partial_a \psi_L + \cdots}$,\         
${\lambda_{Li}(\psi)  = }$                               \\
${ i \xi^I (\psi_L \sigma^I)_i 
+ \kappa^2 \xi^I \gamma^a \psi_{Ri} \bar\psi_R \sigma^I 
\partial_a \psi_L + \cdots}$, 
${A_a(\psi) = - {1 \over 2} \kappa \xi^I \left( 
\bar\psi_L \sigma^I \gamma_a \psi_L 
- \bar\psi_R \sigma^I \gamma_a \psi_R \right) +}$    \\
${ \cdots}$,    
$D^I(\psi) = {1 \over \kappa} \xi^I 
- i \kappa \xi^J \left( 
\bar\psi_L \sigma^I \sigma^J \!\!\not\!\partial \psi_L 
- \bar\psi_R \sigma^I \sigma^J \!\!\not\!\partial \psi_R \right)$ $+ \cdots$, etc.  
and the familiar LSUSY transformations on the component fields of the supermultiplet are 
reproduced in terms of the NLSUSY transformations on  $\psi_L^i$ contained. 
When $\xi^1 = \xi^3 = 0$, the supermultiplet is the ordinary vector supermultiplet 
containning the  vector U(1) gauge field as expected. 
It is remarkable that the compact group SU(2), though global in flat space so far, emerges 
in Riemann-flat spacetime of N=2 SGM (\ref{vaactex}).  \\
This shows that the vacuum of SGM has rich structures manageable, which is favourable to SGM scenario.
\section{Cosmology of NLSUSY GR}
Now we survey the cosmological implications of NLSUSY GR (or SGM from the composite viewpoints). 
SGM spacetime is unstable and spontaneously breaks down to Riemann spacetime 
with superon(massless NG fermion) matter, which is the birth of the universe 
by the quantum effect in advance of so called the  inflation and/or the big bang. 
The variation of (\ref{SGMR}) with respect to  ${e^{a}}_{\mu}$ gives 
the equation of motion for ${e^{a}}_{\mu}$ recasted as follows 
\begin{equation}
R_{\mu\nu}(e)-{1 \over 2}g_{\mu\nu}R(e)=
{8{\pi}G \over c^{4}} \{ \tilde T_{\mu\nu}(e,{\psi})-g_{\mu\nu}{c^{4}\Lambda \over 8{\pi}G} \}, 
\label{SGMEQ}
\end{equation}
where $\tilde T_{\mu\nu}(e,\psi)$ is the stress-energy-momentum tensor of superon(NG fermion) matter 
including the gravitational interactions. 
Note that ${c^{4}\Lambda \over 8{\pi}G}$ can be interpreted as 
{\it the (negative) energy density of empty spacetime}, 
i.e. {\it the dark energy density ${{\rho}_{D}}$}. 
(The sign is fixed by the sign of the kinetic term of VA action.) 
While, on tangent spacetime, the low energy theorem of the particle physics 
gives the following superon(massless NG fermion matter)-vacuum coupling 
\begin{equation}
<{\psi^{L}}_{\alpha}(q) \vert {J^{M\mu}}_{\beta} \vert 0> = 
i\sqrt{c^{4}\Lambda \over 8{\pi}G}(\gamma^{\mu})_{\alpha\beta}\delta^{LM} e^{iqx}+ \cdots,  
\label{LETH}
\end{equation}
where ${{J^{M\mu}}= i\sqrt{c^{4}\Lambda \over 8{\pi}G} \gamma^{\mu}\psi^{M} + \cdots}$      
is the conserved supercurrent obtained by applying the Noether theorem 
to NLSUSY VA action\cite{ks3}        
and $\sqrt{c^{4}\Lambda \over 8{\pi}G}$ is {\it the coupling constant $g_{sv}$ 
of superon with the vacuum}. 
Further we have seen in the preceding section that the right hand side of (\ref{SGMEQ}) for N=2, 
which is essentially N=2 NLSUSY VA action, is equivalent to the broken LSUSY action (\ref{lact}) 
with the vacuum expectation value of the auxiliary field giving the SUSY breaking mass 
$M_{SUSY}$ to a component field of the (massless) LSUSY supermultiplet 
$<D> \sim \sqrt{c^{4}\Lambda \over 8{\pi}G} \sim {M_{SUSY}}^{2}$ . 
We find  NL SUSY(SGM) scenario gives interesting relations among the mysterious quantities 
of the cosmology and the low energy particle physics, i.e., 
$\rho_{D} \sim {c^{4}\Lambda \over 8{\pi}G} \sim 
(fundamental \ length \ of \ SGM \ spacetime)^{-4}$ $\sim <D>^{2}$ \
and \
$g_{sv} \sim \sqrt{c^{4}\Lambda \over 8{\pi}G} \sim <D>$. 
It is natural  to suppose that among the LSUSY supermultiplet the neutrino $\lambda(x)$  
(not superon $\psi(x)$), 
which is the stable and the lightest particle, retains mass of the order of the spontaneous SUSY breaking, 
i.e. ${{m_{\nu}}^{2} \sim \sqrt{c^{4}\Lambda \over 8{\pi}G}}$, 
then SGM predicts the observed value ${{\rho^{obs}}_{D}} \sim (10^{-12}GeV)^{4} \sim {m_{\nu}}^{4}$.%
The tiny neutrino mass is the direct evidence of SUSY (breaking), i.e., 
the spontaneous phase transition of SGM spacetime. 
The proton decay imposes stringent constraints on (SUSY)GUTs in general, so the spacetime origin of 
tiny neutrino mass may be  worth being considered. 
NLSUSY GR gives in general $\Lambda \sim {M_{SUSY}}^{2}({M_{SUSY} \over M_{Planck}})^{2}$.  
The large mass scales and the compact (broken) gauge d.o.f. necessary for the realistic 
and interacting broken LSUSY model will appear 
through the linearization of $\tilde T_{\mu\nu}(e,{\psi})$ which contains the mass scale $\Lambda^{-1}$.
NLSUSY GR(or SGM) (\ref{SGM}) can be easily generalized to spacetime with extra  dimensions, 
which allows to consider the unification in terms of the elementary fields.   \\

\section{Discussions}
The geometry of new spacetime is described by SGM action (\ref{SGM}) of {\it vacuum \ EH-type} 
and gives the unified description of nature. 
As proved for EH action of GR\cite{wttn}, the energy of NLSUSY GR action of 
EH-type is  positive (for $\Lambda>0$).  
NLSUSY GR action (\ref{SGM}), $L(w) \sim \Omega(w) + w\Lambda$, on SGM spacetime is unstable  and 
induces {\it the spontaneous (symmetry) breakdown} into EH action 
with NG fermion (massless superon) matter, 
${L(e,\psi) \sim eR(e) + e\Lambda + (\cdots \psi, e \cdots)}$\cite{st1}, 
on ordinary Riemann spacetime, 
for the cuvature-energy of SGM  spacetime is converted into the energy of Riemann spacetime 
and the energy-momentum of superon(matter), i.e. ${w\Omega > eR}$, which may be called {\it Big Decay}. 
As mentioned before SGM action possesses  two different flat spaces: 
SGM-flat ($w{^a}_{\mu}(x) \rightarrow {\delta}{^a}_{\mu}$)  of NLSUSY GR action $L(w)$ 
and Riemann-flat ($e{^a}_{\mu}(x) \rightarrow {\delta}{^a}_{\mu}$) of 
SGM action ${L(e,\psi)}$ allowing the (generalized) NLSUSY VA action. 
This can be regarded as the phase transition of spacetime from SGM to Riemann 
(with massless NG fermion matter).
Also this may be the birth of the present expanding universe, i.e. the big bang and 
the rapid expansion (inflation) of spacetime and matter, which is followed by 
the present accelleration due to $\Lambda$.  
And we think that the birth of the present universe by the {\it spontaneous \ breakdown} 
of SGM spacetime  described  by {\it vacuum} action of EH-type (\ref{SGM}) 
may explain qualitatively the observed critical value$(\sim 1)$  of the energy density 
of the universe.  
It is interesting if SGM could give new insights into the unsolved problems 
of the cosmology, e.g. the origins(real shapes) of the  big bang, inflation, dark energy,
matter-antimatter asymmetry, $\cdots$, etc.   \par
The linearization of NLSUSY GRT(N=1 SGM) action (\ref{SGM}), i.e. the construction of 
the renormalizable and local {\it broken} LSUSY {\it gauge} field theory which is equivalent to (\ref{SGM}), 
is inevitable to derive the SM as the low energy effective theory and 
to test the abovementioned cosmological aspects as well. 
The linearized theory may have the similarity to SUGRA\cite{fvf}\cite{dz1} 
with the spontaneous SUSY breaking\cite{vs}\cite{dz2} but not the same 
up to the supermultiplet structure, 
for the new global NLSUSY transformation (\ref{newsusy}) should be respected throughout. 
Particularly N=10 must be linearized to test the (composite) SGM scenario for (\ref{SGM}). 
The generalization of the systematics found in ref.\cite{ik} to the superspace formalism\cite{wb} of SUGRA  
may be useful. 
By this detour of the linearization we can circumvent the no-go theorem based on the S-matrix arguments\cite{cm}\cite{hls} for the gravitational 
interaction of the  high spin massless elementry gauge field. 
We can expect the dangerous high spin states are  massive in the linearized equivalent theory 
on the true vacuum, for such states are contained in $\tilde T_{\mu\nu}(e,{\psi})$. 
The study of the vacuum structure of SGM action in the broken phase
(i.e. NLSUSY GRT action in Riemann spacetime with matter) is crucial 
for linearizing SGM.            \par
NLSUSY GR with the extra dimensions, which  can be constructed straightforwardly and 
gives another unification framework by regarding the observed particles as elementary, is open.  
In this case there are two mechanisms for relating the structure of spacetime 
to the dynamical d.o.f., 
i.e. by the compactification of Kaluza-Klein type and by the new mechanism, Big Decay, 
presented in SGM.  \par
\newpage

%
\newcommand{\NP}[1]{{\it Nucl.\ Phys.\ }{\bf #1}}
\newcommand{\PL}[1]{{\it Phys.\ Lett.\ }{\bf #1}}
\newcommand{\CMP}[1]{{\it Commun.\ Math.\ Phys.\ }{\bf #1}}
\newcommand{\MPL}[1]{{\it Mod.\ Phys.\ Lett.\ }{\bf #1}}
\newcommand{\IJMP}[1]{{\it Int.\ J. Mod.\ Phys.\ }{\bf #1}}
\newcommand{\PR}[1]{{\it Phys.\ Rev.\ }{\bf #1}}
\newcommand{\PRL}[1]{{\it Phys.\ Rev.\ Lett.\ }{\bf #1}}
\newcommand{\PTP}[1]{{\it Prog.\ Theor.\ Phys.\ }{\bf #1}}
\newcommand{\PTPS}[1]{{\it Prog.\ Theor.\ Phys.\ Suppl.\ }{\bf #1}}
\newcommand{\AP}[1]{{\it Ann.\ Phys.\ }{\bf #1}}

\end{document}